\documentclass{article}

\usepackage[final]{neurips_2019}

\usepackage{bbm}
\usepackage{float}
\usepackage{amsmath}
\usepackage{graphicx}
\usepackage[utf8]{inputenc} 
\usepackage[T1]{fontenc}    
\usepackage{hyperref}       
\usepackage{url}            
\usepackage{booktabs}       
\usepackage{amsfonts}       
\usepackage{nicefrac}       
\usepackage{microtype}      
\usepackage{subfig}
\title{Making Recommender Systems More Knowledgeable: 

A Framework to Incorporate Side Information}

\author{%
    Yukun Jiang \\
   Language Technologies Institute\\
  Carnegie Mellon University \\
   Pittsburgh, PA 15213 \\
  \texttt{yukunj@cs.cmu.edu} \\
  \And
  Leo Guo\\
  Language Technologies Institute\\
  Carnegie Mellon University\\
  Pittsburgh, PA 15213 \\
  \texttt{jiongtig@cs.cmu.edu }\\
  \And
  Xinyi Chen \\
  Language Technologies Institute \\
  Carnegie Mellon University\\
  Pittsburgh, PA 15213 \\
  \texttt{xinyic@cs.cmu.edu} \\
  \And
  Jing Xi Liu \\
 Language Technologies Institute \\
  Carnegie Mellon University \\
  Pittsburgh, PA 15213 \\
  \texttt{jingxil@cs.cmu.edu} \\
}

\begin{document}

\maketitle
\begin{abstract}
    Session-based recommender systems typically focus on using only the triplet $(user\_id, timestamp, item\_id)$ to make predictions of users' next actions. In this paper, we aim to utilize side information to help recommender systems catch patterns and signals otherwise undetectable. Specifically, we propose a general framework for incorporating item-specific side information into the recommender system to enhance its performance without much modification on the original model architecture. Experimental results on several models and datasets prove that with side information, our recommender system outperforms state-of-the-art models by a considerable margin and converges much faster. Additionally, we propose a new type of loss to regularize the attention mechanism used by recommender systems and evaluate its influence on model performance. Furthermore, through analysis, we put forward a few insights on potential further improvements.
\end{abstract}

\section{Introduction}

The recommender system (RS) has been one of the most successful and profitable AI applications in today's business. Due to the rapidly increasing flow of information online, companies are capable of making timely and suitable recommendations to their customers, in eager hope of facilitating the deal. 

However, training an accurate recommender system is not an easy task. Literature often finds that, for various reasons, traditional statistical machine learning models like K-Nearest Neighbor (KNN) could outperform deep learning-based models, even if the latter is dominating other fields like computer vision and natural language processing. Some conjecture that it is because people do not fully utilize side information when building the model. In this project, we aim to explore a deep neural network approach to incorporate side information into recommender systems to make them more knowledgeable.

Current methods such as MK-MSR proposed by \citet{microbehavior} use item attributes, or item knowledge, to obtain side information for modeling the transition pattern between related segments. However, they did not provide any generic approach that could apply to different recommender tasks. This paper aims to establish a general framework that can be adopted by existing models to improve model performance. 
Experimental results on several models and datasets prove that with side information, our recommender system outperforms state-of-the-art models by a considerable margin and converges much faster. Additionally, we propose a new type of loss to regularize the attention mechanism used by recommender systems and evaluate its influence on models' performance.

The main contributions of this paper can be summarized as the following:
\begin{enumerate}
  \item Demonstrate the effectiveness and usefulness of side information in the context of session-based recommender system
  \item Provide a general framework for incorporating side information into common existing session-based recommender systems
  \item Exemplify how model performance can be further improved without too much modification on its architecture
\end{enumerate}

\section{Related works}
Traditional recommender systems such as content-based RS and collaborative filtering-based RS often integrate the entire user history to understand each user's long-term interests in entities. Newer models using the idea of session-based RS with time-sensitive contexts bring more attention to users' short-term preferences. \citet{sessionsurvey} provided a comprehensive overview of the history and state-of-the-art approaches in session-based recommender systems (SBRS). They distinguished current challenges of SBRS, including recognizing arbitrary interactions within timely sessions, as well as effectively understanding the cascaded long-term sequential correlations that diminish progressively through time. This paper also elucidates inter-session dependencies that span long-range sessions.

For session-based systems, there has been an abundance of novel attempts in neural network based design. \citet{repeatnet} introduced a new neural recommendation architecture called RepeatNet, which exploits the fact that in many scenarios, an item is repeatedly consumed by a user over time. They made the neural network aware of such repeated consumption by a ``repeat-or-explore'' mechanism built upon a regular recurrent neural network (RNN) recommender system. 

Another example is the session-based recommendation with graph neural networks (SR-GNN) proposed by \citet{srgnn}. SR-GNN aims to generate precise item embeddings while taking complex transitions of items into account, where sessions will be modeled as graph structures and items will be modeled as nodes, allowing automatic extraction of features with respect to node connections. SR-GNN presents another method for combining the short-term interests and long-term preferences of users. The model does not assume any distinct latent representation of users, instead, it uses embedding of items in each session to compute short-term and long-term session interests. To combat the inconsistent information priority for each session, a soft-attention mechanism is used to prioritize different interests when calculating the embedding vector representing long-term interest of the session. It then uses session embedding containing long-term interest and short-term interest to generate recommendations. 

Side information has also been explored in previous research. In the paper of \citet{microbehavior}, they made attempts to utilize side information as users' operation, such as reading the comments and adding items to the cart. They incorporated this item knowledge into a multi-task learning paradigm, jointly modeling items and users' micro-behaviors. They defined each micro-behavior as a combination of items and their corresponding operation. By constructing the embedding for both users' and items' side information, their recommender system achieved a remarkable performance boost.

\section{Dataset}
Our experiments are based on four main datasets: an e-commerce transaction dataset Diginetica, a user-based music history dataset Last.FM, a movie rating dataset MovieLens, and a grocery store transaction dataset Ta Feng.
\subsection{Dataset preprocessing}
We apply the following preprocessing steps on each of the datasets to generate train and test sets for further experiments. The specific choice of values are mentioned in Section \ref{dataset}. 

\begin{itemize}
    \item[1.] We choose the training target, e.g. item ID, of which we analyze the patterns and make predictions. We also select the side information, e.g. category or subclass ID, which we utilize to improve our model performance.
    \item[2.] We split data into user-based sessions. The session length varies for different datasets. If the dataset is tested in our baseline models, we follow the same choice of session length. Data within a session spans less than a set duration, e.g. 8 hours or 2 weeks, which varies in different datasets. In addition, we filter out sessions with lengths smaller than 2. We further split sessions if they are longer than the session length.
    \item[3.] We shuffle the dataset and split the session dataset into train and test sets.
    
    \item[4.] For a session like $[x_1, x_2, x_3, \dots, x_n]$, we generate it into $n-1$ lines of data, having $[x_1, x_2, \dots, x_i, 0, 0, \dots, 0] $ being the input and $[x_{i+1}]$ being the target output, for $i \in \{1, \dots, n-1\}$. The length of each input is the predefined session length, and we pad zeros after the $i$-th item. That is to say, the data are generated in the way that given the preceding information within a session, we aim to predict the next item in that session.
    \item[5.] To utilize the side information, we create a mapping of item ID to its side information, e.g. category ID. For each line of the train/test data, we substitute the original item ID into its side information, e.g. category ID, to obtain a complementary dataset. 
    
\end{itemize}

\subsection{Dataset description} \label{dataset}
\textbf{Diginetica}\footnote{\url{https://competitions.codalab.org/competitions/11161\#learn\_the\_details-data2}} This dataset contains user session data extracted from an e-commerce search engine logs that can be used to build session-based recommender systems. We use the same script as that of RepeatNet to preprocess Diginetica in order to replicate the baseline results. In addition to the preprocessing steps mentioned above, we also filter out items that appear less than 5 times. \citet{repeatnet} evaluated RepeatNet's performance on this dataset. However, the RepeatNet model only used basic information like timestamp and product ID to make predictions. We incorporate side information like the category of the product to improve the performance of RepeatNet.

\textbf{Last.FM}\footnote{\url{http://ocelma.net/MusicRecommendationDataset/lastfm-360K.html}} This dataset contains <user, timestamp, artist, song> tuples collected from Last.FM API, which represents the whole listening habits of nearly 1,000 users. For the preprocessing, we select artist ID as the training target and set the session length to be 50, while the session duration is 8 hours. \citet{repeatnet} evaluated RepeatNet's performance on this dataset. This dataset is only used for baseline since it does not contain enough side information.


\textbf{MovieLens}\footnote{\url{https://grouplens.org/datasets/movielens/}} This dataset is a popular movie rating dataset, and the commonly used features are user ID, item ID, timestamp, and rating. For the preprocessing, we select movie ID as the training target and set the session length to be 50, while the session duration is one day. In addition to the preprocessing steps mentioned above, we also filter out items that appear less than 5 times. The MovieLens dataset also contains features like genres of movies, which can be used to train embeddings for movies. We believe such embeddings can be informative and are able to boost the performance of recommender system models.

\textbf{Ta Feng}\footnote{\url{https://www.kaggle.com/chiranjivdas09/ta-feng-grocery-dataset}} This dataset contains transaction data from a Chinese grocery store. Ta Feng includes features such as transaction\_date, customer\_id, product\_id, product\_subclass, etc. For the preprocessing, we select product ID as the training target and set the session length to be 40, while the session duration is 2 weeks. For Ta Feng, we use product subclass as side information to improve the performance.  

\subsection{Side information}

The choice of side information is highly relevant to the effectiveness of the extended model. As most recommender system related datasets contain very few extra information other than item ID, timestamp, and user ID, the most common and accessible side information is often tags or subclass information. In this paper, we only experimented with side information being categorical values, and they represent the higher subclass or category of the targets.

\section{Baseline selection}
We plan to extend the aforementioned deep learning model, the RepeatNet \citep{repeatnet}, and incorporate side information into the current recommender system settings. Therefore, we select RepeatNet as one of our baseline models. We performed modification on the preprocessing part of the original RepeatNet provided by \citet{repeatnet} and run the experiments to replicate the results presented in the paper and compare its performance to our extended model. 


In addition to RepeatNet, we also apply the side information framework to the session-based recommendation with graph neural networks (SR-GNN) proposed by \citet{srgnn} and select SR-GNN as one of our baseline models. The reason is that SR-GNN used Graph Neural Network as encoder, and we want to confirm that our proposed framework works well on various kinds of architecture.

\section{Baseline implementation}

We run experiments on the four selected datasets using both RepeatNet and SR-GNN. 

\subsection{RepeatNet}

\begin{figure}[h!]
\centering
    \includegraphics[width=0.9\textwidth]{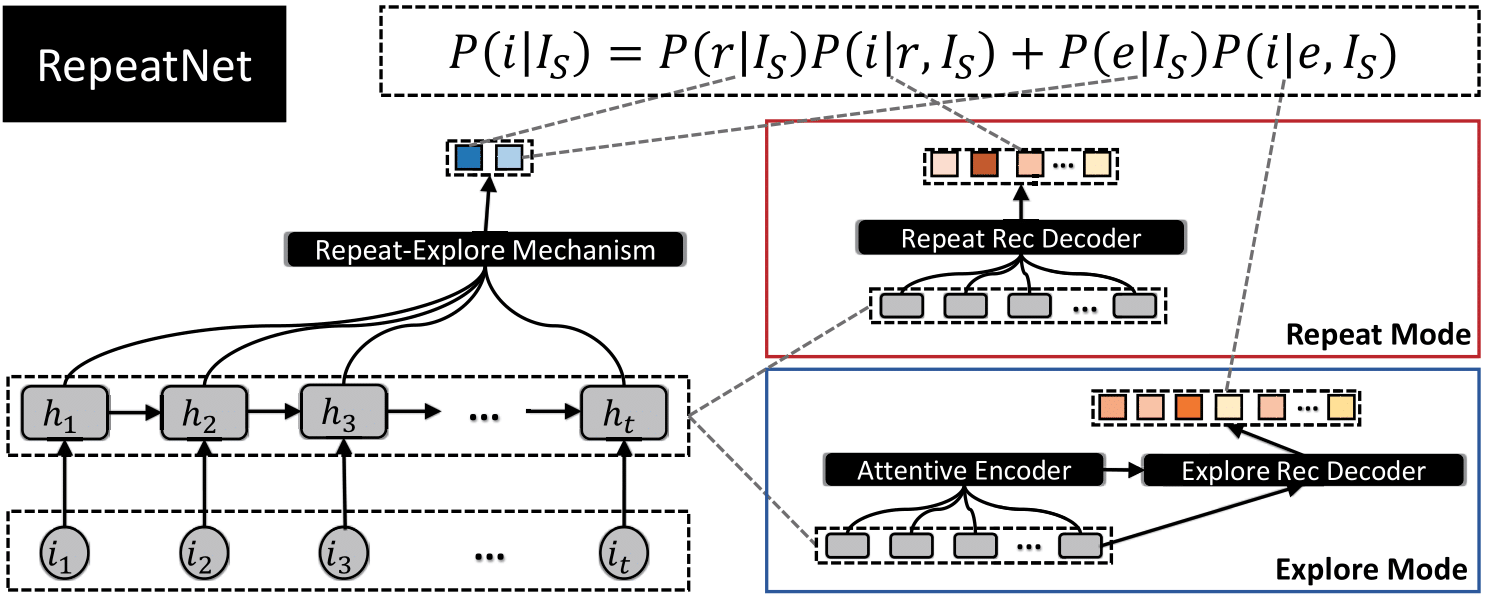}
    \caption{RepeatNet model. This picture is taken from the RepeatNet paper. $\forall j \in \{1, \cdots, t\}, i_j$ represents the item at timestamp $j$, and $h_j$ represents the embedding vector of $i_j$ extracted from a GRU that handles items}
\end{figure}
Among the four datasets, Diginetica and Last.FM are tested by \citet{repeatnet} in the RepeatNet paper. Explanations on the original structure of RepeatNet can be found in Appendix \ref{appendix-repeatnet}. A detailed model description of the extended RepeatNet is presented in Section \ref{repeatnet-description}. We did not simply implement the original RepeatNet architecture. Instead, we modified RepeatNet's explore mode decoder such that the score of each item is computed from the dot product between the item embedding and the vector representing the entire session. In that way, we will be able to incorporate side information when calculating score for each item in our experiments evaluating the effectiveness of side information. As a starting point, we are able to replicate and verify the result of RepeatNet on Last.FM. This means that our modification doesn't negatively influence RepeatNet's behavior. We did not try to replicate the result of RepeatNet on Diginetica, since \citet{repeatnet} used an incorrect method to preprocess Diginetica, as they did not sort items in each session according to timestamp. The result is shown below in Table~\ref{repeatnet-replication}.

\begin{table}[H]
\centering
\begin{tabular}{|l|l|l|}
\hline
Dataset: Last.FM   & \textbf{RepeatNet (original)} & \textbf{RepeatNet (replicate)} \\ \hline
\textbf{Recall@10} & 24.18                         & 24.27                          \\ \hline
\textbf{Recall@20} & 32.38                         & 32.41                          \\ \hline
\textbf{Mrr@10}    & 11.46                         & 11.80                          \\ \hline
\textbf{Mrr@20}    & 12.03                         & 12.36                          \\ \hline
\end{tabular}
\caption{RepeatNet Replication Result} \label{repeatnet-replication}
\end{table}

\subsection{Session-based Recommendation with Graph Neural Networks (SR-GNN)}
\begin{figure}[H]
\centering
    \makebox[\textwidth][c]{%
  \includegraphics[width=1.1\textwidth]{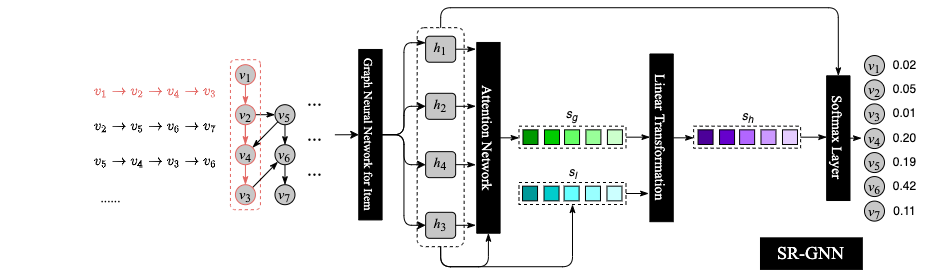}%
}
    \caption{SR-GNN model. This picture is a modification of the original SR-GNN paper. $v$'s represent items in a session and $h$'s represents the embedding vectors of $v$'s extracted from a Graph Neural Network that handles items}
\end{figure}

 Explanations on the original structure of SR-GNN can be found in Appendix \ref{appendix-srgnn}. A detailed model description of the extended SR-GNN is presented in Section \ref{srgnn-description}. Diginetica is tested by \citet{srgnn} in the SR-GNN paper. We are able to replicate the result of SR-GNN on Diginetica, and the result is shown below in Table~\ref{srgnn-replication}.

\begin{table}[H]
\centering
\begin{tabular}{|l|l|l|}
\hline
Dataset: Diginetica   & \textbf{SR-GNN (original)} & \textbf{SR-GNN (replicate)} \\ \hline
\textbf{Recall@20} & 50.73                         & 50.54                          \\ \hline
\textbf{MRR@20}    & 17.59                         & 17.48                         \\ \hline
\end{tabular}
\caption{SR-GNN Replication Result} \label{srgnn-replication}
\end{table}


\section{Algorithm description}

In this section, we propose a method that supports side information incorporation built upon RepeatNet \citep{repeatnet} and SR-GNN \citep{srgnn}. A brief summary of the idea and architecture of RepeatNet and SR-GNN is given in Appendix \ref{appendix-repeatnet} and \ref{appendix-srgnn}. Additionally, we believe that the attention mechanism of existing architectures can be further regularized, so we also propose an additional loss term to regularize the attention.

\subsection{General framework to incorporate side information}
The basic RepeatNet architecture and SR-GNN architecture do not provide a way to incorporate side information of items in each session, which can be useful and provide additional repeating patterns that may boost the performance of the model, even when the items in each session do not repeat. For example, each movie in the MovieLens dataset has genre information, even though people may not watch a film twice within a period of time, they may tend to focus on movies in the same genre within a fixed period of time.

Thus, given an available session $I_S = \{i_1, i_2, \cdots,i_j, \cdots,  i_t\}$, we focus on its corresponding side information $Side_S = \{s_1, s_2, \cdots,s_j, \cdots,  s_t\}$ that has the same length as $I_S$ provided by the dataset, and incorporate such side information into the model. Note that we only consider $\textbf{categorical}$ variables, such as genres of a movie in the MovieLens dataset and the category of the item in the Diginetica dataset when encoding side information. 

As mentioned, we propose a general framework to incorporate and encode side information into any existing session-based recommender system that uses a deep neural network architecture. The most straightforward way to achieve our goal is to directly utilize the encoder of any model we want to modify that encodes items in each session. Our method can be expressed as follows: 
\begin{enumerate}
\item Make a copy of the item encoder that encodes items in each session to encode side information of items in each session.
\item Concatenate the item embedding produced by the item encoder with the corresponding side information embedding produced by the side information encoder. At each timestamp, we will concatenate the item embedding with the side information embedding at the same timestamp.
\item Replace the item embedding with the concatenated embedding in remaining parts of the network.
\end{enumerate}

\subsubsection{Side information encoder of RepeatNet}

\begin{figure}[h!]
\centering
    \includegraphics[width=0.9\textwidth]{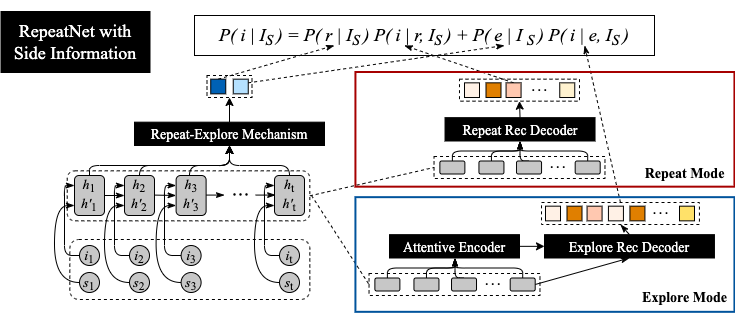}
    \caption{RepeatNet model with side information. $\forall j \in \{1, \cdots, t\}, s_j$ represents the side information at timestamp $j$, and $h'_j$ represents the embedding vector of $s_j$ extracted from a separate GRU that handles side information. The model would concatenate $h_j$ with $h'_j$, and the concatenated vector would be used by the remaining parts of the model}
\end{figure}

\label{repeatnet-description}
In RepeatNet, a Gated Recurrent Unit (GRU) is used to encode session $I_S$ containing items. Specifically, the GRU works as follows:
\begin{align}
    z^i_r &= \sigma(W^i_z[emb(i_r), h^i_{r-1}]) \\ 
    r^i_r &= \sigma(W^i_r[emb(i_r), h^i_{r-1}]) \\
    \Tilde{h^i_r} &= \tanh(W^i_h[emb(i_r), r^i_r\odot h^i_{r-1}])\\
    h^i_r &= (1-z^i_r) \odot h^i_{r-1} + z^i_r \odot \Tilde{h^i_r}
\end{align}

where $W^i_z, W^i_r, W^i_h$, and $emb(i_r)$ are learnable parameters. After this GRU session encoder, a raw available session $I_S = \{i_1, i_2, \cdots, i_t\}$ is encoded into $\{h_1, h_2, \cdots, h_t\}$.

Same as how RepeatNet encodes each session, we will use a GRU to encode side information $Side_S$. $Side_S$ can be either a set of $\textbf{integers}$ or a set of $\textbf{set of integers}$. If $s_j \in Side_S$ is an integer, then it corresponds to $\textbf{one}$ category of the categorical variable, such as the category of the item; if $s_j \in Side_S$ is a set of integer, then it corresponds to $\textbf{multiple}$ categories of the categorical variable, such as the various genres a movie belong to. We will handle these two cases differently. 

$\textbf{Case 1: } Side_S$ is a set of $\textbf{integers}$
\begin{align}
    z^s_r &= \sigma(W^s_z[emb(s_r), h^s_{r-1}]) \\ 
    r^s_r &= \sigma(W^s_r[emb(s_r), h^s_{r-1}]) \\
    \Tilde{h^s_r} &= \tanh(W^s_h[emb(s_r), r^s_r\odot h^s_{r-1}])\\
    h^s_r &= (1-z^s_r) \odot h^s_{r-1} + z^s_r \odot \Tilde{h^s_r}
\end{align}

$\textbf{Case 2: } Side_S$ is a set of $\textbf{set of integers}$
\begin{align}
    z^s_r &= \sigma(W^s_z[\sum_{c\in s_r}\frac{emb(c)}{|s_r|}, h^s_{r-1}]) \\ 
    r^s_r &= \sigma(W^s_r[\sum_{c\in s_r}\frac{emb(c)}{|s_r|}, h^s_{r-1}]) \\
    \Tilde{h^s_r} &= \tanh(W^s_h[\sum_{c\in s_r}\frac{emb(c)}{|s_r|}, r^s_r\odot h^s_{r-1}])\\
    h^s_r &= (1-z^s_r) \odot h^s_{r-1} + z^s_r \odot \Tilde{h^s_r}
\end{align}

where $W^s_z, W^s_r, W^s_h, emb(s_r)$ in case 1, and $emb(c)$ in case 2 are learnable parameters. After this GRU side information encoder, the raw side information $Side_S = \{s_1, s_2, \cdots,  s_t\}$ is encoded into $h^s = \{h^s_1, h^s_2, \cdots, h^s_t\}$.

\subsubsection{Side information encoder of SR-GNN}

\begin{figure}[h!]
\centering
    \makebox[\textwidth][c]{%
  \includegraphics[width=1.1\textwidth]{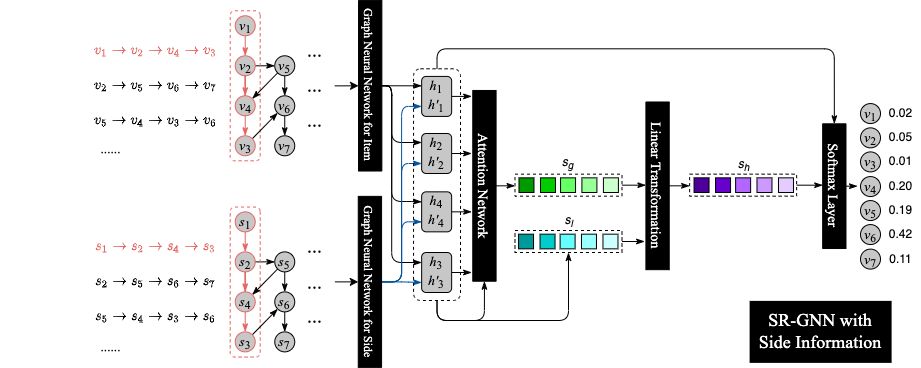}%
}
    \caption{SR-GNN model with side information. $s$'s represent the side information in a session and $h'$'s represents the embedding vectors of $s$'s extracted from a Graph Neural Network that handles side information. The model would concatenate $h_j$ with $h'_j$, and the concatenated vector would be used by the remaining parts of the model}
\end{figure}

\label{srgnn-description}
In SR-GNN, a Graph Neural Network (GNN) is used to encode session $I_S$ containing items. A detailed description of how GNN works can be found in Appendix \ref{appendix-srgnn}. After this GNN item encoder, a raw available session $I_S = \{i_1, i_2, \cdots, i_t\}$ is encoded into $\{h_1, h_2, \cdots, h_t\}$. \\\\
Same as how SR-GNN encodes each session, we will use a GNN to encode side information $Side_S$. However, $Side_S$ can only be a set of $\textbf{integers}$ but not a set of $\textbf{set of integers}$. This means that $s_j \in Side_S$ can only corresponds to $\textbf{one}$ category of the categorical variable, such as the only category of an item, rather than $\textbf{multiple}$ categories of the categorical variable, such as the various genres of a movie. The reason is that GNN uses the edge between two nodes to represent transition between two items or two categories. However, it is hard for GNN to represents transition between two sets of categories since this cannot be simply represented as an edge between two nodes. Thus, we only focus on the case in which $Side_S$ is a set of integers when evaluating how incorporating side information influences the model's behavior, and this means that we will only focus on Diginetica and TaFeng and ignore MovieLens.\\\\
After the GNN side information encoder, the raw side information $Side_S = \{s_1, s_2, \cdots,  s_t\}$ is encoded into $h^s = \{h^s_1, h^s_2, \cdots, h^s_t\}$.

For both RepeatNet and SR-GNN, after obtaining $h^i = \{h^i_1, h^i_2, \cdots, h^i_t\}$ and $h^s = \{h^s_1, h^s_2, \cdots, h^s_t\}$, we generate $h = \{h_1, h_2, \cdots, h_t\}$ based on $h^s$ and $h^i$, where $h_j\in h$ is obtained by concatenating $h^i_j$ and $h^s_j$ together. We then use this new $h$ in remaining parts of the RepeatNet or SR-GNN.

\subsection{Attention loss}
\label{attentionloss}
Ideally, the model should pay more attention to an item in a session if that item is similar to the target item. We believe that we can use the attention mechanism more carefully when calculating the vector representing the entire session in both RepeatNet and SR-GNN. Thus, we here propose the attention loss for items as follows:
\begin{align}
    L_{\text{item attention}}(\theta) = -\frac{1}{|\mathbb{I_S}|}\sum_{I_S\in\mathbb{I_S}}\sum_{r=2}^{|I_S|}\sum_{t=1}^{r-1}\frac{\mathbbm{1}(i_t=i_r)}{|\{j| i_j \in I_S[1:r-1], i_j=i_r\}|}\log(\alpha_t)
\end{align}
where $\mathbb{I_S}$ represents the set containing all sessions and $\alpha_t$ is the attention weight assigned to the item and side information pair at time step $t$. This loss would encourage the attention weights assigned to items that are different from the target item to be 0 and encourage the attention to be uniformly distributed among items that are identical to the target item.\\\\
Additionally, we calculate the attention loss for side information as follows:  
\begin{align}
    L_{\text{side attention}}(\theta) = -\frac{1}{|\mathbb{I_S}|}\sum_{I_S\in\mathbb{I_S}}\sum_{r=2}^{|I_S|}\sum_{t=1}^{r-1}
    \frac{\mathbbm{1}(s_t=s_r)}{|\{j| s_j \in Side_S[1:r-1], s_j=s_r\}|}\log(\alpha_t)
\end{align}
where $Side_S$ represents the side information corresponding to the session $I_S$. The above loss encourages the model to pay more attention to an item in a session if that item and the target item share the same side information, and ignore items that do not share the same side information with the target item.\\\\
We define:
\begin{align}
\text{ItemTarget}(I_S, t, r) = \frac{\mathbbm{1}(i_t=i_r)}{|\{j| i_j \in I_S[1:r-1], i_j=i_r\}|}\\
\text{SideTarget}(Side_S, t, r) = \frac{\mathbbm{1}(s_t=s_r)}{|\{j| s_j \in Side_S[1:r-1], s_j=s_r\}|}
\end{align}
Then, the attention loss we incorporate into the loss term of the model would be:
\begin{align}
    L_{\text{attention}}(\theta) = -\frac{1}{|\mathbb{I_S}|}\sum_{I_S\in\mathbb{I_S}}\sum_{r=2}^{|I_S|}\sum_{t=1}^{r-1}
    \frac{\text{ItemTarget}(I_S, t, r) + \text{SideTarget}(Side_S, t, r)}{2}\log(\alpha_t)
\end{align}
The attention loss defined above encourages the model to pay more attention to an item in a session if that item is identical to the target item, or if that item shares the same side information with the target item. 

\section{Experimental result}
\subsection{Experiment setup}
To achieve fair and comparable experimental results, we follow the model hyperparameter settings as described in the models' original paper (RepeatNet \citep{repeatnet} and SR-GNN \citep{srgnn}). 

Specifically, for RepeatNet, we set item embedding size $s_{\texttt{item\_embedding}} = 100$, side embedding size $s_{\texttt{side\_embedding}} = 100$, hidden size $s_{\texttt{hidden}}=100$, and the item vocab size and side vocab size vary by dataset. We use a batch size of $128$ and an Adam optimizer with initial learning rate $lr = 1e-3$, $\beta_1=0.9$, $\beta_2=0.999$ and the learning rate halves every $5$ epochs.

For SR-GNN, we set we set item embedding size $s_{\texttt{item\_embedding}} = 100$, side embedding size $s_{\texttt{side\_embedding}} = 100$, and the item vocab size and side vocab size vary by dataset. We use a batch size of $100$ and an Adam optimizer with initial learning rate $lr = 1e-3$, $\beta_1=0.9$, $\beta_2=0.999$, weight on $L_2$ penalty being $1e-5$ and the learning rate halves every $3$ epochs with early stopping mechanism.

\subsection{Evaluation metrics}

We mainly use Recall and MRR as our evaluation metrics. We calculate MRR@20 and Recall@20, the percentage of true answer ranking in the top 20 of our predictions. And MRR is the mean reciprocal rank calculated by $\text{MRR} = \frac{1}{N}\sum_{i=1}^N\frac{1}{\text{rank}_i}$. The reciprocal rank is set to 0 if it lies beyond top 20 for MRR@20.

\subsection{Evaluating the effect of incorporating side information}
\subsubsection{Performance} \label{result}

\begin{figure}[h!] 
    \centering
    \subfloat[TaFeng]{%
        \includegraphics[width=0.5\textwidth]{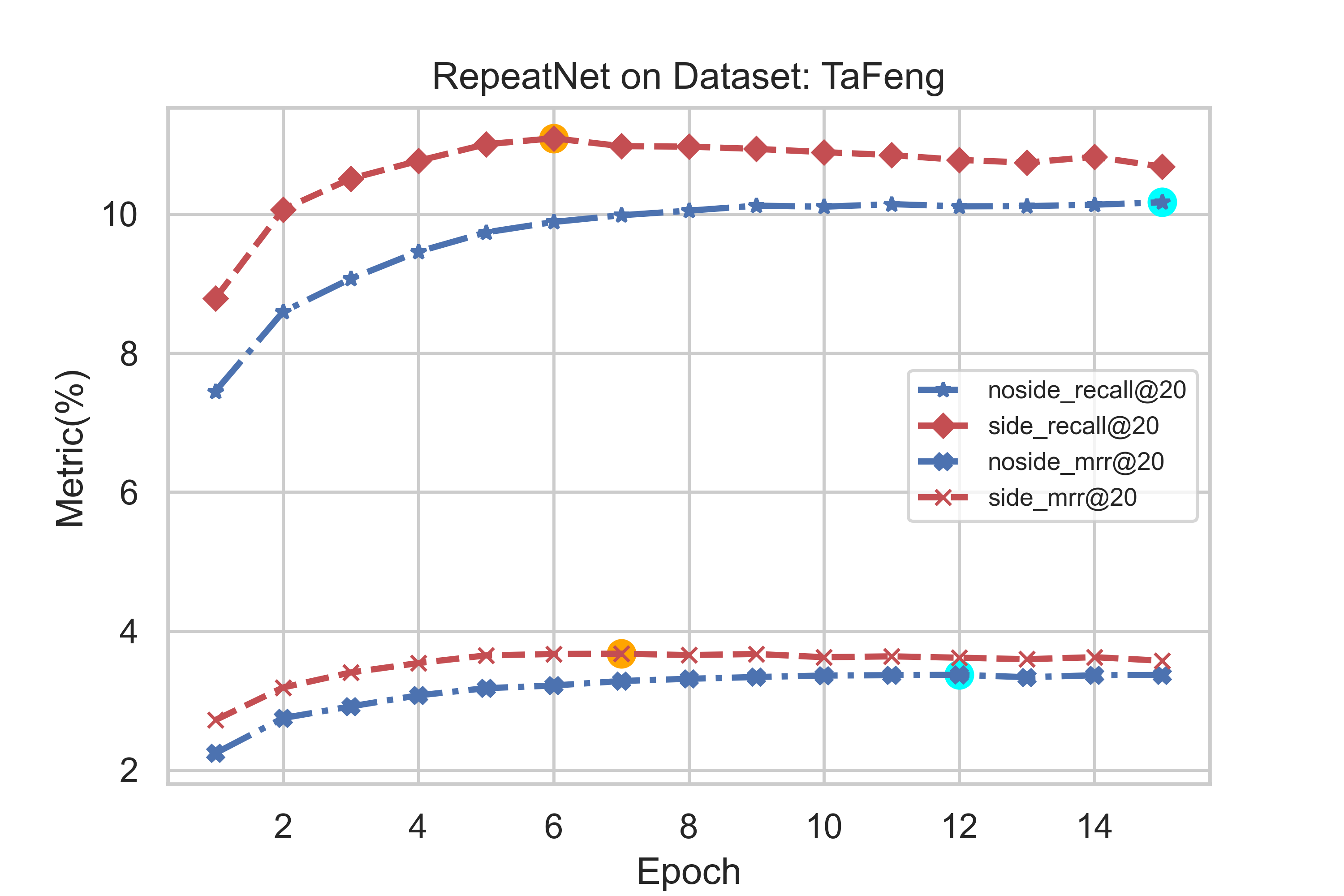}%
        }%
    
    \subfloat[Diginetica]{%
        \includegraphics[width=0.5\textwidth]{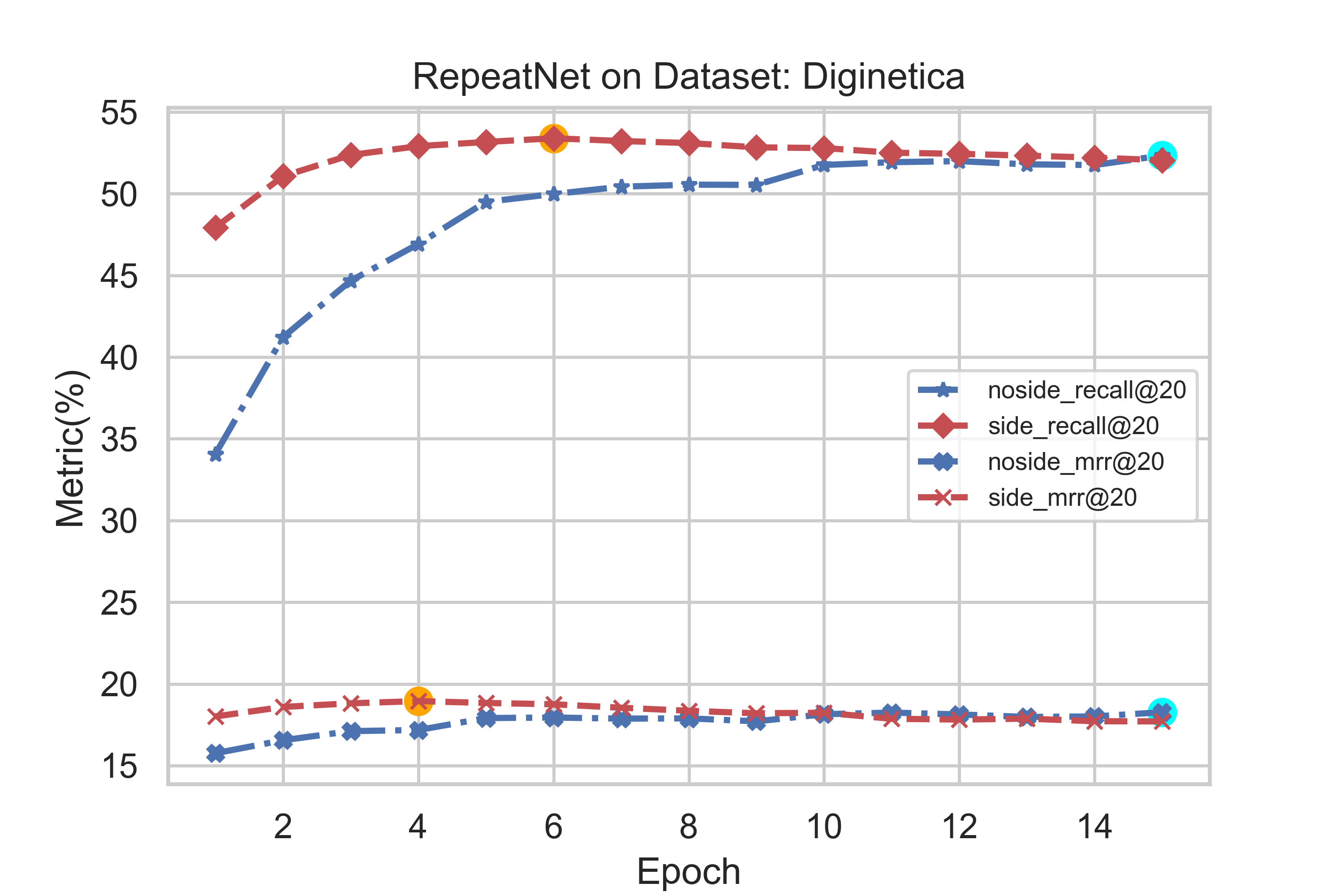}%
        }%
    \hfill%
    \subfloat[MovieLens]{%
        \includegraphics[width=0.5\textwidth]{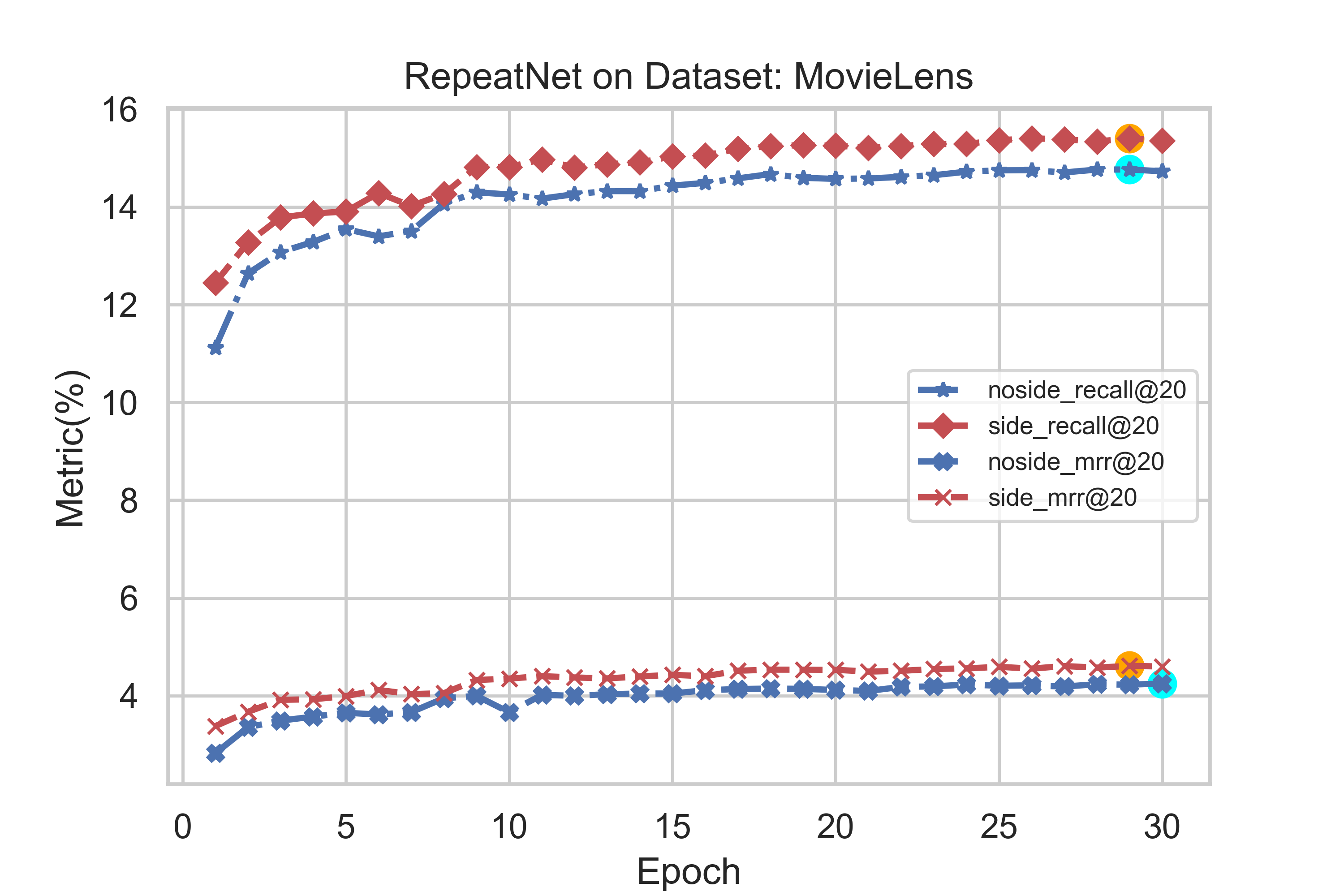}%
        }%
    \caption{RepeatNet Performance \\ Maximum highlighted }
\label{result_figure}
\end{figure}

We summarize our main experimental statistics as below. In addition to Table~\ref{result_table} representing the best results achieved on each dataset in terms of both metrics, we also provide a visualization of the learning process for RepeatNet in Figure~\ref{result_figure}. We will get into detailed analysis in Section \ref{analysis}.

\begin{table}[h!]
\centering
\begin{tabular}{|l|llll|llll|ll|}
\hline
Dataset            & \multicolumn{4}{l|}{\textbf{TaFeng}}                                                                                                                                                         & \multicolumn{4}{l|}{\textbf{Diginetica}}                                                                                                                                                     & \multicolumn{2}{l|}{\textbf{MovieLens}}                                                                                                                                                      \\ \hline
                   Model & \multicolumn{2}{l|}{\textbf{RepeatNet}}                                                                      & \multicolumn{2}{l|}{\textbf{SR-GNN}}                                           & \multicolumn{2}{l|}{\textbf{RepeatNet}}                                                                      & \multicolumn{2}{l|}{\textbf{SR-GNN}}                                           & \multicolumn{2}{l|}{\textbf{RepeatNet}}                                                                                                                \\ \hline
Metrics            & \multicolumn{1}{l|}{\begin{tabular}[c]{@{}l@{}}no\\ side\end{tabular}} & \multicolumn{1}{l|}{side}           & \multicolumn{1}{l|}{\begin{tabular}[c]{@{}l@{}}no\\ side\end{tabular}} & side & \multicolumn{1}{l|}{\begin{tabular}[c]{@{}l@{}}no\\ side\end{tabular}} & \multicolumn{1}{l|}{side}           & \multicolumn{1}{l|}{\begin{tabular}[c]{@{}l@{}}no\\ side\end{tabular}} & side & \multicolumn{1}{l|}{\begin{tabular}[c]{@{}l@{}}no\\ side\end{tabular}} & \multicolumn{1}{l|}{side}           \\ \hline
\textbf{Recall@20} & \multicolumn{1}{l|}{10.17}                                             & \multicolumn{1}{l|}{\textbf{11.09}} & \multicolumn{1}{l|}{9.49}                                                 & \textbf{10.42}    & \multicolumn{1}{l|}{52.33}                                             & \multicolumn{1}{l|}{\textbf{53.39}} & \multicolumn{1}{l|}{50.54}                                                 & \textbf{52.14}    & \multicolumn{1}{l|}{14.77}                                             & \multicolumn{1}{l|}{\textbf{15.40}}    \\ \hline
\textbf{MRR@20}    & \multicolumn{1}{l|}{3.37}                                              & \multicolumn{1}{l|}{\textbf{3.68}}  & \multicolumn{1}{l|}{3.01}                                                 & \textbf{3.21}    & \multicolumn{1}{l|}{18.27}                                             & \multicolumn{1}{l|}{\textbf{18.95}} & \multicolumn{1}{l|}{17.48}                                                 & \textbf{17.64}    & \multicolumn{1}{l|}{4.25}                                              & \multicolumn{1}{l|}{\textbf{4.62}} \\ \hline
\end{tabular}
\caption{performance metrics}
\label{result_table}
\end{table}

\subsubsection{Analysis} \label{analysis}

From Table~\ref{result_table}, we could observe that with side information incorporated, the model performance improves on all metrics by a considerable margin. This applies to all datasets and models we experimented, which exemplifies the universality of our framework for incorporating side information into neural network based recommender system. It is also worth noting that we achieved the state-of-the-art performance on Diginetica dataset. The previous state-of-the-art performance on Diginetica was achieved by 	
NISER+ \citep{gupta2021niser}, and the reported performance was 18.72 in terms of MRR@20 and 53.39 in terms of Recall@20.

Additionally, we notice from the learning process plots that, with side information, RepeatNet converges better and faster if the side information framework is applied. SR-GNN also converges faster and better: without side information, SR-GNN converges in three epochs; with side information, SR-GNN is able to converge in two epochs. This spells a dramatic difference of resources and time invested for real life industry-scale applications.

\begin{table}[h!]
\centering
\begin{tabular}{|l|l|l|l|l|l|l|}
\hline
Dataset: Diginetica   & \textbf{SR-GNN} & \textbf{SR-GNN} & \textbf{SR-GNN} & \textbf{RepeatNet} & \textbf{RepeatNet} & \textbf{RepeatNet} \\ \hline
Embedding Size   & 100 & 200 & 100 & 100 & 200 & 100 \\ \hline
Side information   & Without & Without & With & Without & Without & With \\ \hline
\textbf{Recall@20} & 50.54                       & 49.39          & \textbf{52.14} & 52.33 & 52.00 & \textbf{53.39}                \\ \hline
\textbf{MRR@20}    & 17.48                       & 16.78          & \textbf{17.64} & 18.27 & 18.00 & \textbf{18.95}                \\ \hline
\end{tabular}
\caption{Effect of model's size} \label{model_size}
\end{table}

To confirm that the performance improvement is not resulting from simply increasing the sizes of models, we doubled the size of embedding vectors for both SR-GNN and RepeatNet. This manipulation allows the model without side information to have a same size compared with the model that has a side information encoder. We evaluated the enlarged models on Diginetica dataset, and compared those enlarged models with models with side information encoder. The result is shown in Table~\ref{model_size}. From the table, we can see that increasing the size of model doesn't lead to performance gain, which suggests that our success is not resulting from simply increasing the number of parameters.

\subsection{Evaluating the effect of attention loss}
\subsubsection{Performance} \label{result}

\begin{figure}[h!]
\centering
    \includegraphics[width=0.5\textwidth]{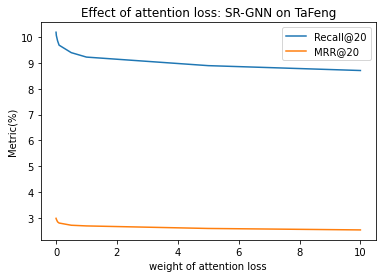}
    \caption{Effect of attention loss}
    \label{attention_loss_plot}
\end{figure}

We experimented with the effectiveness of attention loss by testing SR-GNN model on TaFeng dataset. In particular, we incorporated and tested the attention loss proposed in Section \ref{attentionloss} with various weights. The result is shown in Figure~\ref{attention_loss_plot}.
\subsubsection{Analysis}
From Figure~\ref{attention_loss_plot} we can see that both metrics become worse as we increase the weight of attention loss. This means that the attention loss is not able to boost the performance of the recommender system. We still believe that the model should pay more attention to an item in a session if that item is identical to the target item or that item shares the same side information with the target item. However, our proposed attention loss fails to improve model performance, and we hypothesize that the reason behind is that it is hard to set the correct target for attention used in attention loss, as there is no convincing theory guiding us to set the target. 
%




\section{Conclusion}

In this paper, we present a general framework to incorporate side information into session-based recommender systems using neural network architectures. Our experiments based on RepeatNet and SR-GNN using the Diginetica, MovieLens, and Ta Feng datasets confirm that the use of side information can be easily applied to deep neural network models of various architectures and help boost their performance. Not only does our method achieve better results than prior approaches, including some state-of-the-art models, it also shortens the time taken to converge.

\paragraph{Limitations} There are mainly two limitations of this work. Firstly, the method is only targeting general deep neural networks but may not be applicable to other models like KNN-based models. In addition, the performance of incorporating side information into existing models is highly dependent on the variety and quality of the side information. Secondly, under SR-GNN, the item to side information relationship can only be one-to-one. For example, if the genre of a movie is used as side information, but each movie has multiple genres, SR-GNN is not able to utilize all information except for using only one of them. We believe that these limitations can be insights that lead to promising future directions.

\paragraph{Code} To facilitate reproducibility of our results, we share the code for our experiments at \url{https://github.com/YukunJ/11785-Final-Project-Team15}.

\bibliographystyle{plainnat}
\bibliography{neurips}

\newpage
\appendix

\section{Appendix}
\subsection{RepeatNet Architecture} \label{appendix-repeatnet}


Given an available session $I_S = \{i_1, i_2, \cdots,i_j, \cdots,  i_t\}$ where $i_j$ refers to the $j$-th action (purchasing, browsing, etc.), any session-based recommender system aims to predict the next action through a probabilistic interpretation $\mathbf{P}(i_{t+1}|I_S)$. The RepeatNet utilizes the insights that people tend to repeat their action along the timeline, and design two modes of people's next action: Exploration($e$) or Repeat($r$). Mathematically, it is written
\begin{align}
\mathbf{P}(i_{t+1}|I_S) &= \mathbf{P}(i_{t+1}|I_S, e)\mathbf{P}(e|I_S) + \mathbf{P}(i_{t+1}|I_S, r)\mathbf{P}(r|I_S)
\end{align}

To be able to compute each probabilistic component of the above equation, they designed an attention-based recurrent neural network architecture, mainly in four units, and modified objective function:
\begin{itemize}
    \item Session encoder
    \item Repeat-or-explore mechanism
    \item Repeat mode decoder
    \item Explore mode decoder
    \item Repeat-explore loss
\end{itemize}

\subsubsection{Session encoder}
A Gated Recurrent Unit (\textbf{GRU}) is used to encode the session $I_S$. Specifically, the GRU works as follows:

\begin{align}
    z_r &= \sigma(W_z[emb(i_r), h_{r-1}]) \\ 
    r_r &= \sigma(W_r[emb(i_r), h_{r-1}]) \\
    \Tilde{h_r} &= \tanh(W_h[emb(i_r), r_r\odot h_{r-1}])\\
    h_r &= (1-z_r) \odot h_{r-1} + z_r \odot \Tilde{h_r}
\end{align}

where $W_z, W_r, W_h$, and $emb(i_r)$ are learnable parameters. After this GRU session encoder, a raw available session $I_S = \{i_1, i_2, \cdots, i_t\}$ is encoded into $\{h_1, h_2, \cdots, h_t\}$.

\subsubsection{Repeat-or-explore mechanism}

The next problem is to produce a binary probability of repeat mode and explore mode. To that end, they adopt the attention technique using the last hidden state $h_t$ with each individual hidden state $h_r$. Mathematically, it is 

\begin{align}
    e^{re}_r &= v_{re}\tanh(W_{re}h_t + U_{re}h_r) \\
    \alpha^{re}_r &= \frac{\exp(e^{re}_r)}{\sum_i\exp(e^{re}_i)} \\
    c_{\text{session}}&= \sum_{i} \alpha^{re}_i h_i
\end{align}

where $v_{re}, W_{re}, U_{re}$ are learnable parameters and $c_{\text{session}}$ is the final repeat-or-explore attention vector they attain. The final step is to transform this attention vector is a binary probability through 

\begin{align}
    [\mathbf{P}(r|I_S), \mathbf{P}(e|I_s)] = \text{softmax}(W_{\text{T}} c_{\text{session}})
\end{align}

where again $W_{\text{T}}$ is again a learnable matrix parameter.

\subsubsection{Repeat mode decoder}

The repeat mode decoder essentially computes the probability $\mathbf{P}(i_{t+1}|I_S, r)$ for any action/item on condition that such an action/item has already appeared in the available session. Mathematically, it is computed using attention mechanism as follows:

\begin{align}
    e^{r}_r &= v_r^T\tanh(W_rh_t + U_rh_r)\\
    \mathbf{P}(i|I_S, r) &=\begin{cases} \frac{\sum_i \exp(e^{r}_i)}{\sum_{t}\exp(e^{r}_t)}, \text{if }i\in I_S\\0, \text{if }i\notin I_S\end{cases}\\
\end{align}

where $v_r, W_r, U_r$ are learnable parameters. Notice that the summation in the numerator is because an item $i$ might already repeat multiple times in the available session $I_S$.

\subsubsection{Explore mode decoder}

The explore mode decoder computes the probability of selecting next action/item $i$ that's not in the current available session so far. To capture user's interest in this session, firstly they compute the item-level attention:

\begin{align}
    e^e_r &= v_e^T\tanh(W_eh_t+U_eh_r) \\ 
    \alpha^e_r &= \frac{\exp(e^e_r)}{\sum_t \exp(e^e_t)} \\
    c^e_{\text{session}} &= \sum_t \alpha^e_th_t
\end{align}

where again $v_e, W_e, U_e$ are learnable parameters. Then they concatenate the last hidden state $h_t$ with the attention vector $c^e_{\text{session}}$ into $c_{I_S} = [h_t, c^e_{\text{session}}]$. Finally they use this to compute the explore probability as follows:

\begin{align}
    f_i &= \begin{cases}-\infty, \text{if }i\in I_S\\V_ec_{I_S}, \text{if }i\not\in I_S \end{cases}\\
    \mathbf{P}(i|I_S, e) &= \text{softmax}(f_i)
\end{align}

\subsubsection{Repeat-explore loss}

In training the RepeatNet, the team not only uses the normal negative log-likelihood loss function but also adds an explore-repeat loss to facilitate the learning of when to switch the explore mode or repeat mode. Mathematically, we define the negative log-likelihood loss as

\begin{align}
    L_{\text{normal}}(\theta) = -\frac{1}{|\mathbb{I_S}|}\sum_{I_S\in\mathbb{I_S}}\sum_{r=2}^{|I_S|}\log(\mathbf{P}(i_r|I_S[:r-1]))
\end{align}

where $I_S[:r-1]$ represents the first $r-1$ items of session $I_S$, and the repeat-explore loss as

\begin{align}
    L_{\text{r-e}}(\theta) = -\frac{1}{|\mathbb{I_S}|}\sum_{I_S\in\mathbb{I_S}}\sum_{r=2}^{|I_S|}\mathbbm{1}(i_r\in I_S[:r-1])\log(\mathbf{P}(r|I_S[:r-1])) + \mathbbm{1}(i_r\not\in I_S[:r-1])\log(\mathbf{P}(e|I_S[:r-1])) 
\end{align}
and the final loss function objective is defined as

\begin{align}
    L(\theta) = L_{\text{normal}}(\theta) + L_{\text{r-e}}(\theta)
\end{align}

\subsection{SR-GNN Architecture}
\label{appendix-srgnn}
For SR-GNN, we follow the implementation of \cite{srgnn}. SR-GNN aims to generate precise item embeddings while taking complex transitions of items into account, where sessions will be modeled as graph structures and items will be modeled as nodes, allowing automatic extraction of features with respect to node connections. The model uses an attention network to represent each session's long-term interest, and each session is represented as the combination of long-term interest and current interest. It then uses session embeddings created from latent vectors of items to generate recommendations. Their design can be summarized into 4 components:
\begin{itemize}
    \item Construct session graphs
    \item Learn item embeddings of session graphs
    \item Generate session embeddings
    \item Make recommendations
\end{itemize}

\subsubsection{Construct session graphs}
SR-GNN uses a directed graph $\mathcal{G}$ to model each session, where items are represented as nodes of the graph. User interactions as click or purchases are represented as edges, where edge $(\mathcal{V}_{s,i-1}, \mathcal{V}_{s,i}) \in \mathcal{G}$ $\mathcal{E}_s$ 
means the user clicked item $i-1$ before item $i$ in the current session $s$, such that the transitions of items can be captured by the model. The model normalized the weights of the edges to account for repetitions of items. The weight is calculated by $\displaystyle \frac{\textrm{number occurrence of edge}}{\textrm{out degree of edge’s start node}} $.

\subsubsection{Learn item embeddings of session graphs}
Each item $v \in V$ is embedded into a unified space and the node vector $\mathbf{v} \in \mathbb{R}^d$ represents the latent vector of item $v$ learned through graph neural networks.
For graph $\mathcal{G}_s$ corresponding to session $I_S$, the weight connections of incoming and outgoing edges are represented by matrices $A^{(out)}_s$ and $A^{(in)}_s$, with $A_s$ being the concatenation of these matrices in each session. 
\\\\
For the node $v_{s, i}$ of graph $\mathcal{G}_s$ corresponding to session $I_S$, the update functions are given as follows:
\begin{align}
    a^t_{s,i} &= A_{s, i:} [\mathbf{v}^{t-1}_1, ..., \mathbf{v}^{t-1}_n]^TH+b \\
    z^t_{s,i} &= \sigma(W_z a^t_{s,i} + U_z \mathbf{v}^{t-1}_i) \\ 
    r^t_{s,i} &= \sigma(W_r a^t_{s,i} + U_r \mathbf{v}^{t-1}_i) \\
    \tilde{\mathbf{v}^t_i} &= tanh(W_oa^t_{s, i} + U_o(r^t_{s, i} \odot \mathbf{v}^{t-1}_i)) \\
    \mathbf{v}^t_i &= (1-z^t_{s, i}) \odot \mathbf{v}^{t-1}_i + z^t_{s, i} \odot \tilde{\mathbf{v}^t_i}
\end{align}

Where $A_{s, i:} \in \mathbb{R}^{1 \times 2n}$ is a row in $A_s$ corresponding to node $v_{s, i}$ and $H \in \mathbb{R}^{d \times 2d}$ controls the weight. Information is propagated by equation (38), using the restrictions of A as the model proceeds each session. It extracts latent vectors of adjacent nodes, which are later fed into the network. The update gate $z_{s,i}$ decides what information is retained, and the reset gate $r_{s,i}$ decides what is discarded. The final state combines the hidden states and the current candidate state, which is determined by the previous and current states, and the final node vector is retrieved when all nodes in the session are updated until convergence. 

\subsubsection{Generate session embeddings}
SR-GNN uses node vectors in each session to represent the current interest and the long-term interest of each session. Each session is represented as a long-term-interest embedding vector $\textbf{s}_g$ that is obtained from aggregating embeddings of all items in the session, concatenated with the current-interest embedding vector $\textbf{s}_l$, which is the embedding of the last item in the session. SR-GNN uses a soft-attention mechanism was used to prioritize different interests when calculating the long-term-interest embedding vector. 

\begin{align}
    \alpha_i &= q^T \sigma(W_1\mathbf{v}_n + W_2\mathbf{v}_i + c) \\
    s_g &= \displaystyle\sum^n_{i=1}\alpha_i\mathbf{v}_i
\end{align}

Where $q \in \mathbb{R}^d$ and $W_1, W_2 \in \mathbb{R}^{d \times d}$ control the item embedding vectors' weights. The final hybrid embedding is obtained by a linear transformation of the concatenation of the current-interest embedding vector and long-term-interest embedding vector.

\begin{align}
    s_h = W_3 [s_1; s_g]
\end{align}

\subsubsection{Make recommendations}
A recommendation score $\hat{z}$ is computed using the embedding of each session calculated in the last component by:

\begin{align}
    \hat{z}_i = s^T_h\mathbf{v}_i
\end{align}

The output vector $\hat{y}$ of model is obtained using a soft-max function of the recommendation score:

\begin{align}
    \hat{y} = softmax(\hat{z})
\end{align}

They define their loss function using cross-entropy of the predicted value and the real value:

\begin{align}
    L(\hat{y}) = -\displaystyle\sum^m_{i = 1} y_i log(\hat{y}_i) + (1 - y_i)log(1 - \hat{y}_i)
\end{align}

Where y is a one-hot encoding of the real value of each item.

\end{document}